\documentclass{article}

\usepackage{cite}
\usepackage{amsmath,amssymb,amsfonts}
\usepackage{algorithmic}
\usepackage{url}
\usepackage[ruled]{algorithm2e} % Write pseudo-code algorithms
\SetKwRepeat{Dowhile}{do}{while} % Do-while function
\usepackage{graphicx}
\usepackage{textcomp}
\usepackage{xcolor}
\usepackage{lipsum} % For dummy text
\usepackage{tabularx}
\usepackage{adjustbox}
\usepackage{multirow}
\usepackage{rotating}
\usepackage{makecell}
\usepackage{comment}
\PassOptionsToPackage{hyphens}{url}\usepackage[hidelinks]{hyperref}
\usepackage{subcaption}
\usepackage{changepage}
\usepackage{geometry}
\title{GeoFINDR: Practical Approach to Verify Cloud Instances Geolocation in Multicloud \\
\thanks{This work was supported by the France 2030 ANR Project ANR-23-PECL-0009 TRUSTINCloudS.
}}

\author{Saïd Ider\\
\textit{Samovar, Télécom SudParis, } \\
\textit{Institut Polytechnique de Paris}\\
Palaiseau, France \\
said.ider@telecom-sudparis.eu
\and
Maryline Laurent\\
\textit{Samovar, Télécom SudParis, } \\
\textit{Institut Polytechnique de Paris}\\
Palaiseau, France \\
0000-0002-7256-3721
}

\date{April 2025}

\begin{document}

\maketitle

\begin{abstract}
In multicloud environments, where legal obligations, technical constraints and economic interests are at stake, it is of interest to stakeholders to be able to locate cloud data or the cloud instance where data are decrypted for processing, making it particularly vulnerable.
This paper proposes an original and practical delay-based approach, called GeoFINDR, to locate a cloud instance, e.g. a Virtual Machine (VM), over the Internet, based on RIPE Atlas landmarks. First, the assumed threat model and assumptions are more realistic than in existing solutions, e.g. VM-scale localization in multicloud environments, a Cloud Service Provider (CSP) lying about the VM's location. Second, the originality of the approach lies in four original ideas: (1) geolocalization is performed from the VM, (2) a Greedy algorithm selects a first set $LM_A$ of distributed audit landmarks in the vicinity of the declared area, (3) a sectorization algorithm identifies a set $LM_S$ of other landmarks with distance delay behavior similar to that of the VM to estimate the sector of the VM, and (4) the estimated location of the VM is calculated as the barycenter position of the $LM_S$ landmarks.
An open source tool is published on GitHub and experiments show that localization accuracy can be as high as $22.1km$, under unfavorable conditions where the CSP lies about the location of the VM.
\end{abstract}

\section{Introduction} \label{chap:Introduction}
Even if customers specify in their SLAs that they want their data and data processing to be located in data centers in certain regions or cities, it is difficult for them to be sure that their SLA is being met, because most cloud services operate like a complex black box. Verifying the actual location of data and workloads from the declared cloud region is virtually impossible and never 100\% reliable anyway, even when expensive means are deployed.

Geolocalization of data and physical resources on the Internet is very useful \cite{data localization impacts cyber and cloud protection}\cite{importance data center locations} for customers and also for cloud service providers (CSP) themselves, in meeting their obligations to customers in multicloud environments and avoiding penalties. (1) For legal and sovereignty reasons, there is a need to have guarantees where the externalized data is stored to benefit from GDPR protection \cite{gdpr}, to avoid the CLOUD Act, which allows the US to collect data from US companies or data stored in US cloud servers. (2) For availability and resiliency reasons, customers have an interest in ensuring that their virtual resources are replicated in different geographic locations to avoid, for example, losing all their data in the event of a fire, as happened in 2021, resulting in OVH paying customers at least 400,000€ in compensation\footnote{\href{https://www.transatlantic-lawyer.com/ovh-must-pay-more-than-400000-e-after-a-fire-destroyed-its-data-centers-why-this-decision-is-important-for-hosting-providers-hosting-eu-personal-data/}{https://www.transatlantic-lawyer.com/ovh-must-pay-more-than-400000-e-after-a-fire-destroyed-its-data-centers-why-this-decision-is-important-for-hosting-providers-hosting-eu-personal-data/}}. (3) To combat mistrust of CSP, it is important to provide guarantees on the location of virtual resources.

This paper proposes a new approach called GeoFINDR, to verify the physical location of cloud instances, e.g. a VM, over the Internet. GeoFINDR is practical in many aspects. It supports VM-scale localization, which has a higher granularity than the cloud server loacalization, which is provided by most of the geolocation solutions of the literature. Localizing at VM scale is interesting as it matches the requirements of multicloud environments and the VM is known as a vulnerable instance, since it is there that data is decrypted before being processed. An open source tool is available on GitHub \footnote{\href{https://github.com/diasredi/GeoFINDR}{https://github.com/diasredi/GeoFINDR}
}, and experiments show upto $22.1km$ accuracy with an average of $22.6km$, which are very good results compared to other solutions. Note that these measures were obtained under unfavorable conditions where the CSP lies about the location of the VM.

The paper is organized as follows. After positioning our GeoFINDR solution against related works on delay-based solutions in Section \ref{related works}, Section \ref{proposed model} presents our GeoFINDR approach, from the threat model and assumptions to the improvements brought to the area of delay-based solutions. Section \ref{implementation} describes the implementation details, including the resulting pseudocode. Section \ref{experimentation} presents the experiments performed with the methodology and fully explains the achieved results, before concluding in Section \ref{chap:Conclusion}.

\section{Related Works on Delay-based Solutions}\label{related works}
Delay-based solutions rely on multilateration to locate an entity, without CSP intervention, and derive the position of an entity from a set of landmarks with known positions and a method for calculating the distance between the entity and the landmarks. Table \ref{tab:similar works} summarizes the main features of the existing solutions. 

CBDG (Constraint-Based Data Geolocation) \cite{cbdg} is the first software implementation for locating data over the Internet. CBDG uses PlanetLab servers as landmarks and calculates the \textbf{Distance-Delay Relation (DDR)} by linear regression of the ping RTT (Round Trip Time) between all landmarks. All landmarks then ping the server to measure RTT and derive a distance that is used to define areas where the server might be located. We emphasize that attempting to locate data is pointless, since data is duplicated anyway due to the high-availability architecture of the cloud and can circulate by physical means. Therefore, we consider the issue of data location outside the scope of our study and instead focus our efforts on verifying the location of a cloud instance, which is relevant because it is where encrypted data is processed.

In IGOD \cite{igod}, the authors improve CBDG DDR computation, overall performance, and accuracy by implementing several refinements, such as using file transfers instead of pings, and computing a minimum regression instead of an average regression. In addition, IGOD introduces a while loop that selects three dispersed landmarks in an area where the server is located, which gets smaller until it reaches a small enough satisfactory size. They showed that dispersed and closer landmarks gave more accurate results than more distant and randomly chosen landmarks. In doing so, they significantly reduced the number of landmarks needed to estimate the server's position compared to CBDG with greater accuracy.

In SPLITTER \cite{splitter}, the authors use a "double ring triangulation" to reduce the area in which the server could be located, and they manage to reduce the uncertainty of an estimate, but they do not address common issues like the risk of zero intersections.

In VLOC \cite{vloc}, the approach to server localization is reversed from previous solutions. Instead of performing the audit from outside the server, they propose a method for each VM to perform the audit internally using available landmarks. VLOC use web pages as landmarks with approximate positions due to low reliability of IP databases \cite{schopman}. DDR is defined by average delays of measurements with landmarks combined with a trust coefficient to be specified.

The advantage of internal audit, as proposed by VLOC and our GeoFINDR approach, is the ability to localize at the cloud instance level, while the external audit limits location to a single server instance or cloud PROXY. Therefore, internal audit allows for better scalability and adaptation to cloud and multicloud requirements by being able to verify the location of each VM instance in the cloud's backend and include in-cloud delays.

Note that the threat model differs from one approach to another, as shown in Table \ref{tab:similar works}. CBDG, IGOD and SPLITTER consider the \textit{economically rational CSP}, i.e. a CSP that has an interest in moving data to cheaper regions to operate and has no interest in having multiple copies of the same data for cost reasons. This model is not realistic because a CSP can classically run multiple cloud instances, or make copies of them, and the CSP can move them arbitrarily, which is referred to as the "dishonest CSP" model in the table. The difference between VLOC and GeoFINDR threat model is that GeoFINDR considers a realistc dishonest CSP which can use dark fibers, i.e. a faster private network between sites.

Note also that neither paper explains how to handle cases where the distance estimates have no intersections. We believe that for unstable measurements such as network delays and inaccurate DDR, the probability of not finding an intersection by multilateration is high and needs to be addressed to have an automated deployable solution. GeoFINDR provides a solution to this problem thanks to the barycenter calculation (cf. Section \ref{estimation position}).

\section{Our GeoFINDR Approach }\label{proposed model}
GeoFINDR is based on a realistic threat model and assumptions, which are presented in Section \ref{hypothesis}. 
GeoFINDR proposes to improve the 3 key stages of delay-based geolocation solutions, namely: (1) the Greedy landmarks selection algorithm (cf. Section \ref{choice-of-lm}), which selects audit landmarks $LM_A$ in the vicinity of the declared area, 
(2) the sectorization (cf. Section \ref{distance-delay relation}) based on a new definition of the Distance-Delay Relation (DDR) and the selection of landmarks $LM_S$ with similar distance delay behavior as the VM,
and (3) the estimation of the position with barycenter calculation (cf. Section \ref{estimation position}).

\begin{table*}[htbp]
\caption{Comparative Delay-based Solutions including our GeoFINDR Approach}
    %\centering
    \begin{center}
   %\resizebox{\textwidth}{!}{% >{\raggedright\arraybackslash}X
    %\begin{tabular}{|c|c|c|c|c|c|c|}
    %\begin{tabularx}{\textwidth}{|c|c|c|c|c|c|c|}
    \begin{adjustwidth}{-0.65cm}{}
    \begin{tabularx}{18.33cm}{|c|c|c|c|c|c|c|}
        \hline
        \textbf{Method} & \textbf{Audit type} & \textbf{Delay$^{\mathrm{a}}$} & \textbf{Landmarks} & \textbf{Threat model} & \textbf{Contributions} & \textbf{Results}\\
        \hline
        CBDG \cite{cbdg} & external & internet & PlanetLab & \makecell{economically\\rational CSP} & \makecell{reference\\method} & \makecell{median 166km \\90\% $<$626km}\\
        \hline
        IGOD \cite{igod} & external & total & PlanetLab &  \makecell{economically\\rational CSP} &  \makecell{reduce needed\\LM amount} & avg 88.5km\\
        % &  &  & & reduce needed LM amount & avg 88.5km\\
        \hline
%        DPVGeo \cite{dpvgeo} & identify region change & PoR & cloud servers & economically & detection of CSP falsification & 78.6\% of clouds respect SLA\\
%       &  &  &  &  rational CSP &  & \\
        %\hline
        SPLITTER \cite{splitter} & external & total & PlanetLab & \makecell{economically\\rational CSP} & \makecell{double ring\\triangulation} & \makecell{best 30km\\avg 139km}\\
        \hline
        VLOC \cite{vloc} & internal & total & Web sites & dishonest CSP & VM geolocation & best 150km\\
        \hline
        \textbf{GeoFINDR}  & \textbf{internal} & \textbf{total} & \textbf{RIPE Atlas} & \makecell{\textbf{realistic}\\ \textbf{dishonest CSP} }&   \textbf{VM geolocation}& \makecell{\textbf{best 22.1km}\\ \textbf{avg 22.6km}$^{\mathrm{a}}$}\\
    \hline
    %\end{tabular}%}
    \end{tabularx}
    \end{adjustwidth}
    \end{center}    
    \footnotesize{$^{\mathrm{a}}$Total delay means Internet and in-cloud delay \\
    $^{\mathrm{b}}$Results of the tolerance experiment (Section \ref{experimentation}) with parameters $tolerance=100$, $zone\_size=1000$, $NB\_LM=15$, $interval\_percent=15\%$}
    \label{tab:similar works}

\end{table*}

\subsection{Realistic Threat Model and Assumptions}\label{hypothesis}
The considered threat model corresponds to the realistic dishonest CSP model presented in Section \ref{related works}. That is, the CSP is inclined to lie about the declared location of the VM and has the following capabilities:

%We consider the CSP as the main threat to the geolocation of data in the cloud. Hence, we propose a more realistic threat model that includes CSP's real capabilities over data and VMs:
\begin{enumerate}
    \item Total control and unlimited access to data and hardware in the cloud,
    \item High performance architecture (dark fiber, cutting edge hardware, etc.),
    \item Multicloud architecture (large scale cloud, subcontractor, etc.),
%    \item CSP is inclined to lie on the real position of VM than declared %errors or falsification of proofs (i.e., false position declaration, use of a canal DC1 → DC2 → internet).
    %tampering: ralentir les ping ; block LM
\end{enumerate}

GeoFINDR also relies on the following assumptions to verify the location of the VM/cloud instance: 
\begin{enumerate}
    \item There is another machine with similar performance (DDR) to the VM that is connected to the Internet,
    \item Delay measured from two Internet machines has a low variance in time,
    \item Two machines on the same network with similar configuration and hardware will get the same RTT when pinging a third machine,
    \item The cloud instance is a IaaS or PaaS that is connected to the Internet and able to run GeoFINDR,
    \item The cloud instance is able to send and receive packets to/from the landmarks.    
\end{enumerate}

\subsection{Greedy Audit Landmark Selection Algorithm} \label{choice-of-lm}
As reported in previous studies \cite{igod}, dispersed and closer landmarks have better overall accuracy than more distant and undispersed landmarks. Therefore, there is a need for the maximum amount of available landmarks, with multiple positions around the world to maximize their density and dispersion. 
For technical purposes, landmarks need verified positions, continuous calculation of delays between them, transparent measurement data to derive DDR, a limited set of hardware configurations to minimize the variance of response delays between different landmarks (so position is more important), and a public IP address without PROXY to guarantee that we communicate only with the desired landmark.
For all of these reasons, the landmarks of RIPE Atlas were chosen (see Section \ref{implementation:ripe atlas}). 

% audit LM = dispered landmarks in an initial area. If not enough, augment area size.
Once the size of $LM_A$ is specified as the $NB\_LM$ parameter (cf. Table \ref{tab:variables}), which is the amount of audit landmarks required, we need to select distributed landmarks in an area of size $zone\_size$ around the estimated location of the VM, as shown in Figure \ref{fig:geofindr run A}. $NB\_LM$ and $zone\_size$ are input parameters to GeoFINDR. We start with a low value of $zone\_size$ to reduce communication time and get better accuracy. If there are not enough landmarks in the area, we increase $zone\_size$ until there are enough.

Landmark latitude and longitude coordinates can be described as points placed on a space, the Earth. The most dispersed points in a scatterplot are those that maximize the sum of their distances from each other. 

To efficiently select dispersed landmarks, we implemented an original Greedy algorithm (see Algorithm \ref{alg:dispoints}) that selects dispersed points \ref{alg:dispoints} called Dispoints with a $\theta(n\times m)$ complexity that is new to our knowledge. In addition to the great circle distance \cite{zucconi great circle distance}, used to compute the physical distance between two landmarks, we obtain an efficient way to obtain a deterministic set of dispersed landmarks, as shown in Figure \ref{fig:geofindr run B}.
%How it works: from a set of points, select n points. For all points, if a point A's sum of it's distance with each previously selected points is greater than the sum of the closest selected point C's distance with others selected points, replace the closest selected point C with the point A. End loop, return selected points.

\begin{algorithm}
    \caption{Greedy Algorithm to Select $n$ Dispersed Points - Dispoints Code Available at \href{https://github.com/diasredi/dispoints}{https://github.com/diasredi/dispoints}%TBA (removed for the blind review process) 
    }
    %ML ajouter le GitHub où trouver l'algorithme https://github.com/diasredi/dispoints
    \label{alg:dispoints}

    \KwData{list of $n$ points with coordinates, function $distance$ to calculate a distance between 2 points, positive integer $m$}
    \KwResult{list of $m$ dispersed points $result$}
    $selected \gets$ list of $m$ points\; %selected = list des m premiers points
    \For{each point $p$ \textbf{in} $points$}{

        $nearest\_point \gets $ point from $selected$ that minimizes its distance with $p$\; %nearest_point = min de distances
        
        delete $nearest\_point$ from $selected$\; %supprimer le nearest point de selected
        \eIf{sum of distances between $p$ and all $selected$ points $>$ sum of distances between $nearest\_point$ and all $selected$ points
        }{
            Append $p$ to $selected$\;
        }{
            Append $nearest\_point$ to $selected$\;
        }% si la somme des distances entre p et les selected > somme des distances entre nearest_point et selected 
        % alors ajouter à selected p
        % sinon ajouter à selected nearest_point (reviens comme avant)
       
    }
     $result \gets selected$\;
\end{algorithm}

\subsection{DDR-based Sectorization Principle}\label{distance-delay relation}

The distance-delay relation (DDR) is the key component to get accurate estimations of the positions. If the DDR is inaccurate, the error will propagate to the position estimation and produce very large deviations.

\begin{figure}[htbp]
\centerline{\includegraphics[width=8.5cm]{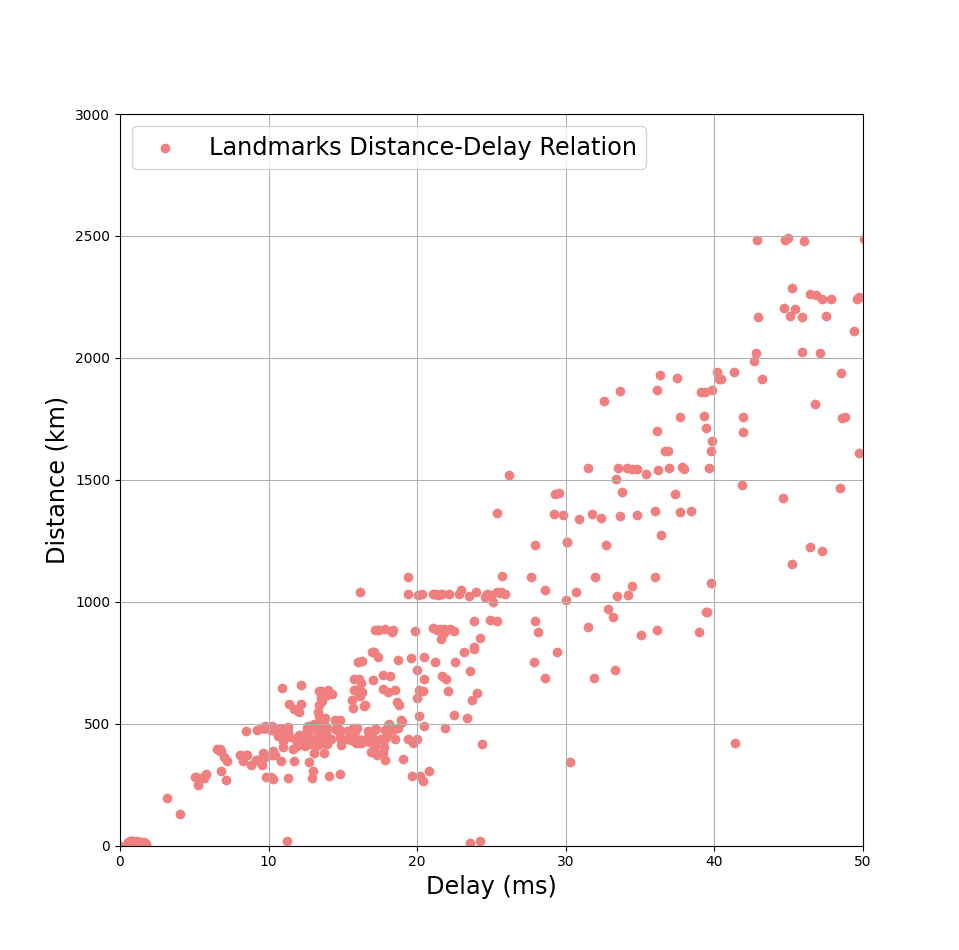}}
\caption{Example of the Distance-delay Relations Distribution between a Landmark and all available Landmarks under 50ms}
\label{fig:ddr plot}
\end{figure}

Previous works generalize the DDR either with theoretical formulas or empirically by measuring delays between each landmark. However, for some landmarks, the generalized DDR is incorrect and leads to large deviations. Looking at the anchor measurements from RIPE Atlas, even with similar hardware and configurations, we find that distances can vary too much for a given measured RTT. For example, in Figure \ref{fig:ddr plot}, for an RTT of about 20ms, the distances vary from 250 to 1100 km. This makes it unreasonable to construct the DDR by regression with acceptable error rates. Defining a unique DDR for each landmark or machine tested, results in too many errors and inaccuracies, demonstrating the need for a dynamic approach to define DDR based on each landmark.

GeoFINDR's intuition regarding DDR is the following one: since distances can vary so much for a given delay, and the "true" distance value for a delay is likely to be close to one of the measurements (to one or more landmarks), the question becomes "Which landmark has the closest DDR to my VM?". To find the most "similar" landmark, we have to make several measurements with other landmarks to infer it. A landmark with the exact same configuration and placement as our VM will have very close delays and distances with other landmarks.

\begin{figure}[htbp]
\centerline{\includegraphics[width=8.5cm]{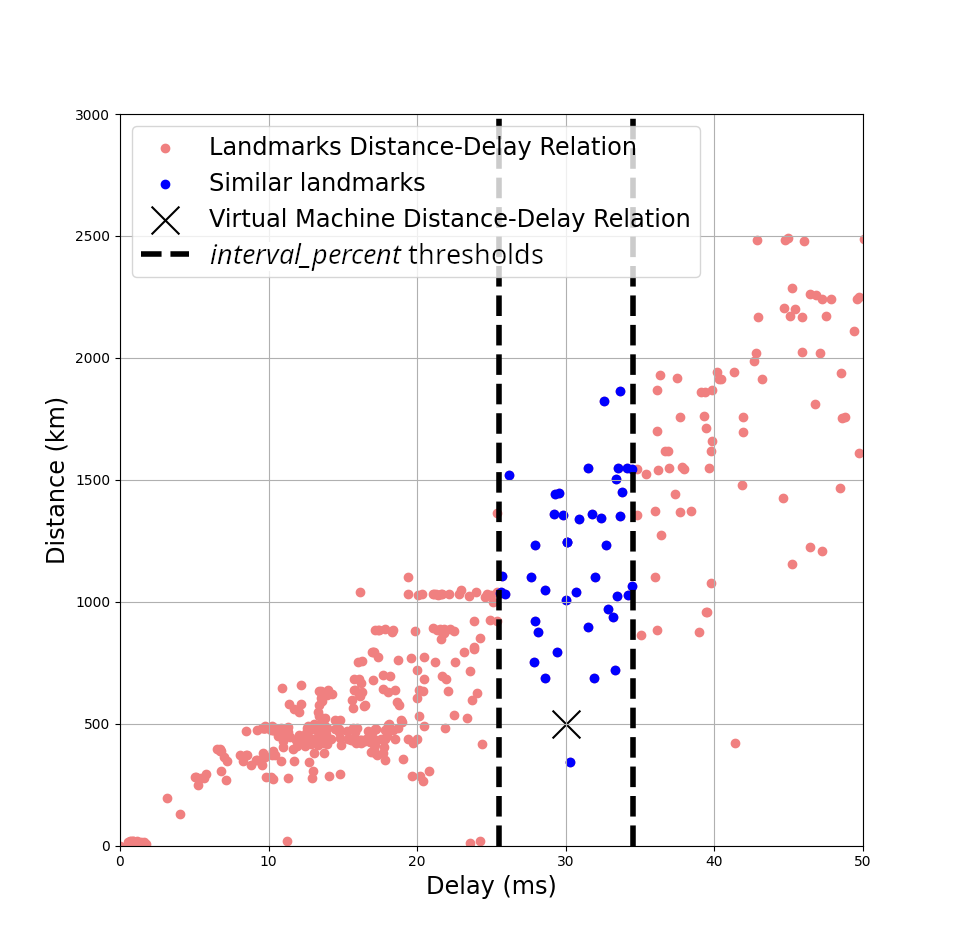}}
\caption{Example of GeoFINDR Operation to Determine a Set of $LM_S$ using the Distance-Delay Relations between a $LM_A$ and other Landmarks, based on the Measured Delay of 30ms with the VM}
\label{fig:ddr plot interval}
\end{figure}

GeoFINDR consists of finding a set $LM_S$ of similar landmarks that have close distance-delay behavior to our VM to infer a sector where the VM might be located, as shown in Figure \ref{fig:geofindr run C}. Similar landmarks are found by measuring delays with the defined set $LM_A$ of audit landmarks, then landmarks with delays equal to our VM +/- $interval\_percent$ are selected, as shown in Figure \ref{fig:ddr plot interval}. The similar landmarks, i.e. landmarks with close DDR, are the landmarks with the most occurrences between all measurements.

\subsection{Estimation of the Position}\label{estimation position}

\begin{figure*}
\begin{subfigure}[t]{.45\textwidth}
  \centering
  \includegraphics[width=.97\linewidth]{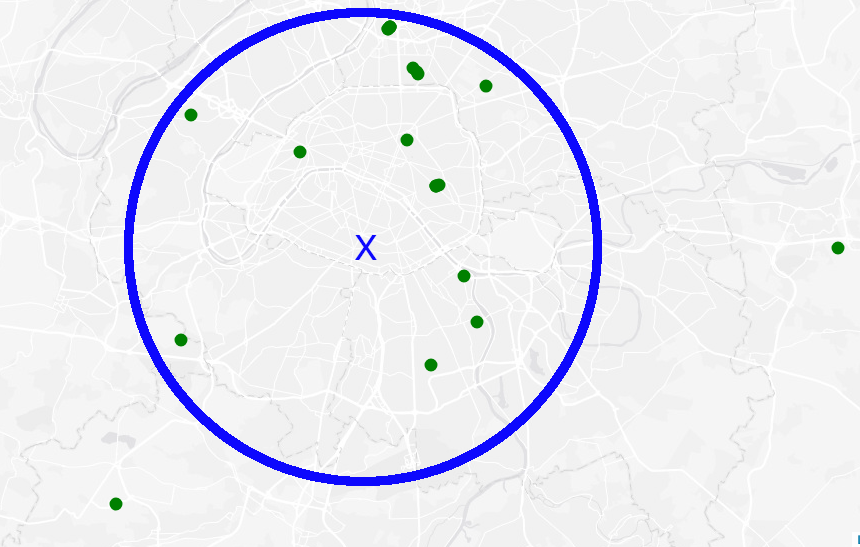}
  \subcaption{Available Landmarks in the Area of $zone\_size$ km around the Declared Position (X)\\}
  \label{fig:geofindr run A}
\end{subfigure}
\hfill
\begin{subfigure}[t]{.45\textwidth}
  \centering
  \includegraphics[width=.97\linewidth]{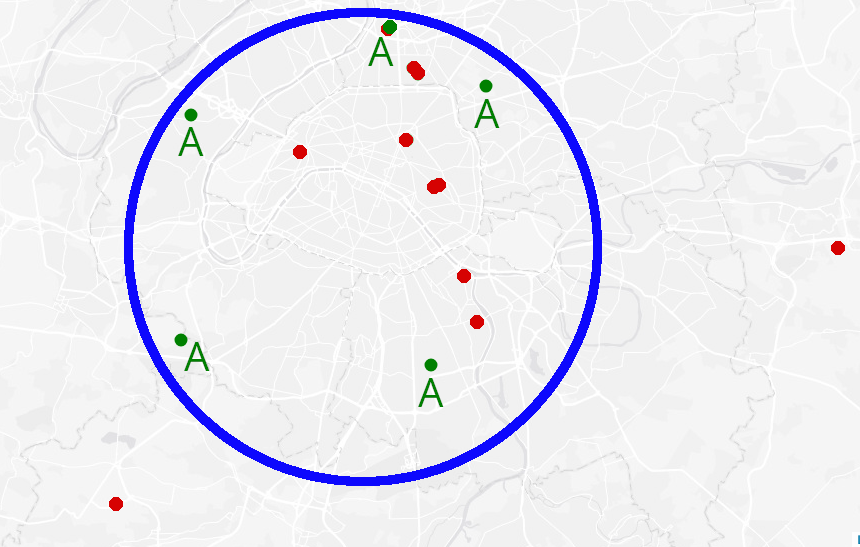}
  \subcaption{Selection of Dispersed $LM_A$ Subset (A)}
  \label{fig:geofindr run B}
\end{subfigure}
%\hfill

\begin{subfigure}[t]{.45\textwidth}
  \centering
  \includegraphics[width=.97\linewidth]{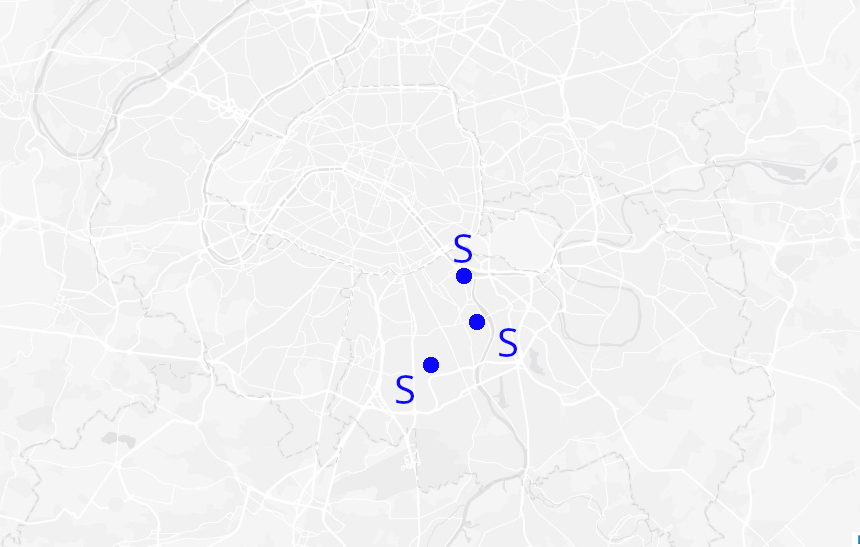}
  \subcaption{Selection of $LM_S$ (S)}
  \label{fig:geofindr run C}
\end{subfigure}
\hfill
\begin{subfigure}[t]{.45\textwidth}
  \centering
  \includegraphics[width=.97\linewidth]{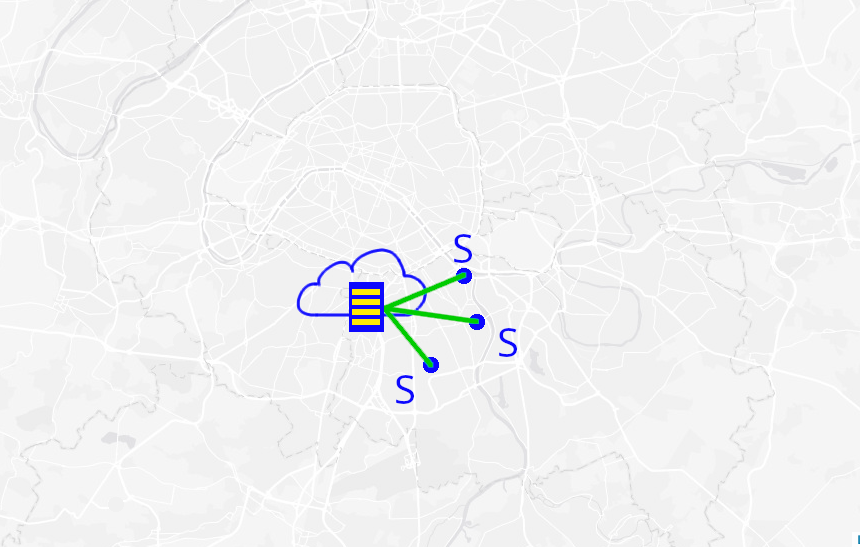}
  \subcaption{Estimation of VM's Position using Barycenter Calculation with $LM_S$}
  \label{fig:geofindr run D}
\end{subfigure}
\caption{Illustration of GeoFINDR Operations}
\label{fig:geofindr run}
\end{figure*}

Ideally, similar landmarks of set $LM_S$ (cf. Section \ref{distance-delay relation}) should be close to the true position of our VM if there are enough landmarks nearby. However, similar landmarks may be too far away to define a satisfactory position. Therefore, we estimate the position of the VM by calculating the barycenter of all landmarks within $LM_S$ weighted by the delays measured between them and our VM.
To solve the zero intersection problem identified in \ref{related works}, we first consider multilateration as an optimization problem instead of an intersection \cite{zucconi multilateration}. The goal is to find the position that minimizes its squared distances with the circles drawn by the estimated distances around the landmarks. By dividing the square root of the squared distances, we obtain an indicator over the estimation's quality, the squared mean root error (SMRE), zero being a perfect intersection.
To obtain the barycenter, we normalize the measured delays between each $LM_S$ and multiply them by the lowest distance that minimizes the given SMRE in previous step. The lowest distance obtained then allows us to deduce the optimal position, the barycenter, as shown in Figure \ref{fig:geofindr run D}.

\section{GeoFINDR Implementation}\label{implementation}
%dans implem: choix de ripe atlas, python etc.. 
This section includes details on the choice of the RIPE Atlas landmark network, the resulting pseudocode, and the specifics of the Python implementation.  
The reader is invited to refer to the notations in Table \ref{Notations} to better understand the pseudocode and related explanations. 

\subsection{RIPE Atlas Landmarks}\label{implementation:ripe atlas}
Most implementations have used PlanetLab servers as landmarks. The problem is the relatively low density of available server locations on it's network, the high amount of different hardware configurations and the difficulty to implement permanent measurements.

For deployment in real use cases, we recommend to deploy a custom set of landmarks to have more control over each audit step. For this study, we used the RIPE Atlas network to demonstrate the validity of our approach.

We chose RIPE Atlas anchors as landmarks because of their limited hardware configurations, public IP addresses, worldwide presence with about 850 available anchors, open source permanent real-time measurements between anchors, and verified geographic location \cite{ripe anchors}.

\subsection{Pseudocode }\label{high level algorithm}

The pseudocode of our GeoFINDR approach is provided in Algorithm \ref{alg:high level alg}, which refers to the three key steps described in Section \ref{proposed model}, namely, the Greedy algorithm for selecting dispersed audit landmarks ($LM_A$), the DDR-based sectorization estimate to measure delays with audit landmarks and infer similar landmarks ($LM_S$), and the VM location estimate with the barycenter calculation. These steps are contained in a do-while loop that ends when the distance between the last sector coordinates and the previous one is less than $tolerance$. If both are close enough to end the loop, we consider the result to be "converged". It is necessary to define a threshold to end the loop because the estimates depend on measured delays, a very volatile value that can produce non-significant differences between two results that we consider satisfactory. If the solution doesn't converge, the main do-while loop allows us to redo the estimates by moving the initial search point to correct the estimate and thus obtain more accurate results.

\SetKwComment{Comment}{/* }{ */}

\begin{algorithm}
    \caption{GeoFINDR Pseudocode - Full Code available at URL \href{https://github.com/diasredi/GeoFINDR}{https://github.com/diasredi/GeoFINDR}%TBA (removed for the blind review process)
    }
    \label{alg:high level alg}
    
    \KwData{$declared\_position$, $zone\_size$, $tolerance$, $interval\_percent$, $NB\_LM \geq 1$}
    \KwResult{$estimated\_position$}
    %$declared\_position \gets (declared\_latitude;declared\_longitude$)\;
    
    import $landmarks$\;
    $sector\_coordinates \gets declared\_position$\;
    
    \Dowhile{Distance($initial\_position, sector\_coordinates$) $\geq tolerance$ }{
        $initial\_position \gets sector\_coordinates$\;

        $near\_landmarks \gets landmarks$ \textbf{where} $distance(initial\_position,landmark) < zone\_size$\;

        $LM_A \gets Dispoints(near\_landmarks, Distance(),NB\_LM)$\;

        $similars\gets \emptyset$\;
    
        \For{$lm$ \textbf{in} $LM_A$}{
            $measured\_delay \gets $ Ping($lm.ip$)\;
            add in $similars$ landmarks with measured RTT from $lm$ in $[measured\_delay-interval\_percent, measured\_delay+interval\_percent]$\;
            
        }

        $LM_S\gets$ landmarks with maximum occurrences in $similars$\;
    
        $measured\_delays \gets \emptyset$\;
        \For{$lm$ \textbf{in} $LM_S$}{
            add in $measured\_delays$ Ping($lm.ip$).RTT\;
        }
        $distance\_scale \gets$ Normalize($measured\_delays$)\;
        $sector\_coordinates \gets$ Barycenter($similar\_landmarks, distance\_scale$)\;

    }
    $estimated\_position \gets sector\_coordinates$\;
\end{algorithm}

\subsection{Python Implementation}\label{implementation:python}
We use Python 3 to implement our model due to its large presence in scientific fields, high-level programming, wide compatibility across systems, and high number of libraries, including the RIPE Atlas library \cite{cousteau}.

Measurements for the audit uses the ICMP ping protocol in IPv4, using the same amount of pings and payload as the RIPE Atlas measurements (3 pings and 64 bytes of payload by default). RTT includes in-cloud and Internet delay. The in-cloud delay is obtained by measuring the ping delay to the server's public IP address or PROXY and can be used to detect potential fraud, as we expect a low in-cloud delay for an honestly declared position.

To estimate the location of the VM, we used the great circle distance function provided in \cite{zucconi great circle distance} to calculate the distance between two coordinates in kilometers using their real latitude and longitude.

\section{Experimentations}\label{experimentation}

The full methodology used in the experiments is described in Section \ref{experimentation:protocol}, followed by several experiments presented in Sections \ref{exp:setup params} and \ref{exp:landmark_meshing}.

\subsection{Methodology}\label{experimentation:protocol}
To test our method, we use a 64bit 8-core CPU @ 1.60GHz, 32Gb RAM, Linux kernel version 6.12.11-200.fc41.x86\_64 (64-bit) computer connected via a 1Gb/s link at the Telecom SudParis campus in Evry (25km south of Paris). This machine plays the role of the VM in a datacenter that needs to be located in our experiment. The advantage is that we have the environment completely under control, and we know its position and the landmarks around. The aim is to have a fixed VM and fictitiously modify the surrounding environment (for example, the number of landmarks available in the area). 

Let's not lose sight of the fact that we want to show whether our GeoFINDR solution can reliably 1) detect a lie in the SLA's declared position and 2) find the true region of the datacenter from different declared positions in Europe and the world.

Our GeoFINDR algorithm (Algorithm \ref{alg:high level alg}) uses four setup parameters that users must configure: $tolerance$, $zone\_size$, $NB\_LM$ and $interval\_percent$ (cf. Table \ref{Notations}). Each parameter changes the behavior of our solution and more or less affects the accuracy and the time to complete the audit.

To measure the impact of each parameter, we are able to calculate two accuracy metrics - the accuracy of estimation with $distance\_real\_declared$, and the ability to detect a lie with $distance\_estimation\_declared$ - and two performance metrics: $nb\_iterations$ and $audit\_time$.
Finally, our solution measures $loopbackRTT$ and $proxyRTT$ throughout the audit to give the user an estimation of the cloud instance's process time and in-cloud delay, respectively, by pinging the cloud's public IP address.
All these features are defined in Table \ref{Notations} with their associated implications.

The following experiments are conducted to better understand the behavior of GeoFINDR according to: 
\begin{enumerate}
    \item \textbf{the Setup Parameters} (cf. Section \ref{exp:setup params}): after selecting an arbitrarily set of parameter values that give us satisfactory results for each declared position, i.e. $tolerance=100$, $zone\_size=200$, $NB\_LM=15$ and $interval\_percent=15$, we modify each one of them individually to observe its effect on the accuracy of the estimates (i.e. low average $distance\_estimation\_real$) and the performance (i.e. average time $audit\_time$ to complete the audit) of GeoFINDR, from a set of real and false declared positions.
% we change the setup parameters $zone\_size$, $interval\_percent$, $tolerance$ and $NB\_LM$ to observe their influence on the accuracy and performance of GeoFINDR, from a set of real and false declared positions.
    \item \textbf{the Landmark Meshing} (cf. Section \ref{exp:landmark_meshing}): after defining an optimal set of parameters from the previous experiment, i.e. $tolerance=100$, $zone\_size=1000$, $NB\_LM=16$ and $interval\_percent=35$, we consider two mesh configurations simulating two landmark configurations: an area dense in landmarks (Paris area), and a less populated area by removing available landmarks in the Paris area to simulate a dead zone (rural or depopulated in landmarks area) to observe the influence of the landmark distribution on the  estimates.
    %we consider XXX different meshing configurations to simulate the end of a land zone (IrelandXXXX  scenario) or a more or less populated landmarks: Africa scenario for depopulated landmarks, XXX for a little more populated landmarks and XXX for locally depopulated landmarks.  
\end{enumerate}

To measure the ability of GeoFINDR to detect a false declaration and find the true position of the cloud instance, we change for both experiments, the declared position from the true position to known capital cities around the world, i.e. 24 positions: Evry (true position), Amsterdam, Tirana, Andorra la Vella, Helsinki, Paris, Berlin, Budapest, Reykjavik, Rome, Warsaw, Lisbon, Moscow, Madrid, Stockholm, Bern, London, Torshavn, Tokyo, Melbourne, Dubai, Buenos Aires, Los Angeles, and Kinshasa.

We consider an estimation satisfactory if the distance between the estimated position and the true position is less than 50km. This is the size of the Paris area. In the case of a true declaration, we expect the estimated position of the VM to be close to the declared position. In the case of a low accuracy, the solution will still be able to detect a lie in the declared position if the distance between the estimated and declared positions is greater than the $tolerance$, the opposite would indicate that the method cannot detect a misleading declaration and is therefore vulnerable.

Note that in real use, the \textit{true} position of the VM can't be inferred, hence the need to measure GeoFINDR's ability to recover accurate estimates in a controlled environment.

\begin{table*}[htbp]
\caption{GeoFINDR Setup Parameters, Notations and Metrics} \label{Notations}
    \centering
%    \begin{center}
    %\begin{tabularx}{\textwidth}{|>{\centering\arraybackslash}X|>{\centering\arraybackslash}X|>{\centering\arraybackslash}X|>{\centering\arraybackslash}X|}
    \begin{adjustwidth}{-0.65cm}{}
    \begin{tabularx}{18.665cm}{|c|c|>{\raggedright\arraybackslash}X|>{\raggedright\arraybackslash}X|}
    
%    \begin{tabular}{|c|c|c|c|}
        \hline
        \textbf{Type}&\textbf{Notation} & \centering\textbf{Function} & \makecell{\centering\textbf{Implication}} \\
        \hline
        \multirow{4}{*}{\makecell{setup\\parameters}}&$tolerance$ & 
        distance threshold between the last two estimated positions, used to terminate the audit & too low and the audit is longer; too high and the audit is inaccurate \\ \cline{2-4}
        %\hline
        &$zone\_size$ & size of area to select $LM_A$, augmented if not enough landmarks available & higher values allow measurements with additional landmarks, low impact on accuracy\\ \cline{2-4}
        %\hline
        &$NB\_LM$ & size of the set $LM_A$ & lower size increases the size of $LM_S$ \\ \cline{2-4}
        %\hline
        &$interval\_percent$ & tolerated deviation between measured delays to make up the set $LM_S$ & higher values increases the size of $LM_S$ \\ 
        %\hline
        \hline
        \makecell{accuracy\\metrics} & $distance\_A\_B$ & distance in kilometers between positions A and B (declared, real or estimated) & $distance\_real\_declared$ measures the extent of the lie, $distance\_real\_estimated$ is the accuracy, $distance\_estimated\_declared$ is the lie estimation \\
        \hline
        \multirow{2}{*}{\makecell{performance\\metrics}}&$nb\_iterations$ & number of iterations of the main loop until convergence performed by the algorithm  & higher values means more measurements and estimations \\ \cline{2-4}
        &$audit\_time$ &  audit duration & higher values mean more time to estimate the position of the VM \\ 
        \hline
        \multirow{2}{*}{\makecell{in-cloud\\measurements}}&$loopbackRTT$ & loopback address ping RTT & estimate VM's internal ping processing time, lower means lower VM capacity \\ \cline{2-4}
        
        &$proxyRTT$ & proxy's address ping RTT & estimate in-cloud delay, high value reflects complexity of flow routing within the CSP (e.g. dark fiber)\\ 
        
        %SMRE
        
 %copy& More table copy$^{\mathrm{a}}$& &  \\
%\hline
%\multicolumn{4}{l}{$^{\mathrm{a}}$Sample of a Table footnote.}
    \hline
    \end{tabularx}
    \end{adjustwidth}
    \label{tab:variables}
%    \end{center}
\end{table*}

%\subsection{Setup/testbed}\label{experimentation:testbed}

\subsection{Impact of Setup Parameters} \label{exp:setup params}

\textbf{zone\_size.}
We set the initial $zone\_size$, the size of the area in which $LM_A$ are selected around the initial position, to 1, 50, 100, 250, 500, 750, 1000, 1500, 2000, 3000, 5000, 7500, 10000 and 15000 kilometers, respectively.
Overall, we found that changing the initial $zone\_size$ has a small effect on accuracy, with only 3 out of 336 estimates giving more than 50km from the true position. For lower values of $zone\_size$, $zone\_size$ is gradually increased until there are enough $LM_A$ inside, and higher values don't affect the accuracy, thus showing the ability of GeoFINDR to estimate a sector even with more $LM_A$. However, the effect on the average audit time is up to $78.2s$ for very large values of $zone\_size$ due to higher RTT delays from distant $LM_A$, and the longer time to select the dispersed $LM_A$ due to the higher number of landmarks in the selection area.
We recommend setting the initial $zone\_size$ to the size of the desired audit area to combine both a lower audit time and accurate estimates.

\textbf{interval\_percent. }
We set $interval\_percent$ 
%the percentage deviation to consider a landmark as similar based on the measured RTT with a LMA, 
to go from 5\% to 105\%, with a step of 10\%. Increasing $interval\_percent$ increases the number of similar landmarks $LM_S$ for each measurement. 

\begin{figure}[htbp]
\centerline{\includegraphics[width=8.7cm]{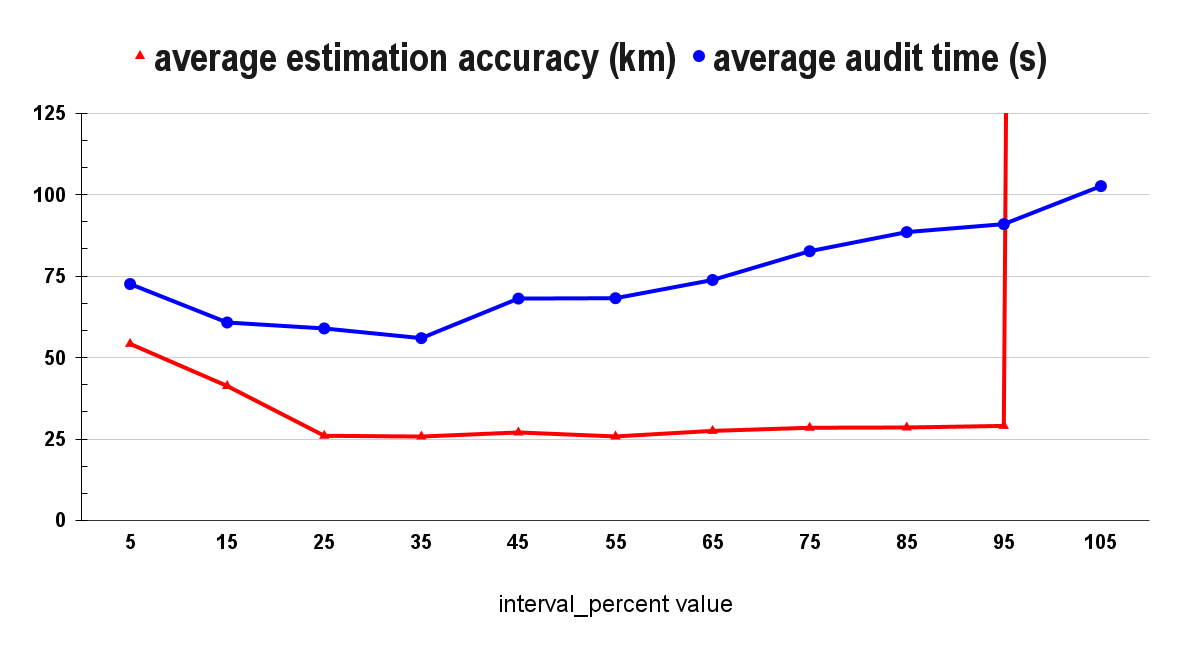}}
\caption{Average Estimation Accuracy and Average Audit Time as a Function of $Interval\_Percent$ ($tolerance=100$, $zone\_size=200$, $NB\_LM=15$)}
\label{fig:interval}
\end{figure}

As shown in Figure \ref{fig:interval}, lower values give less accurate estimates because the $interval\_percent$ is too restrictive and many $LM_A$ do not lead to finding similar landmarks, which doesn't allow the most similar landmarks to be selected. Results are satisfactory from $interval\_percent=15\%$ to $interval\_percent=95\%$, which shows that the selected $LM_S$ are in the desired region. However, increasing the amount of $LM_S$ has limitations. First, the probability of selecting a $LM_S$ outside the true region increases, which can lead to more errors: At $interval\_percent=105\%$, two very distant $LM_S$ are chosen, which ruins the estimation with a $4250.6km$ error on average. Second, more $LM_S$ means more measurements to perform to determine the barycenter, which significantly increases the time required to complete the audit (from an average of $56s$ for $interval\_percent=35\%$ to an average of $102.6s$ for $interval\_percent=105\%$).

In our experiment, the best result and the fastest audit time are obtained with $interval\_percent=35\%$ with an average $dist\_estimation\_real=25.8km$ and an average audit time of $56s$.

\textbf{tolerance. }
We set $tolerance$
%, the distance threshold between the last two estimated positions in kilometers where we consider that GeoFINDR results have converged and thus end the audit, 
to 1, 10, 25, 50, 75, 100, 150, 200, 250, 300, 350, 400, 500, 750, 1000, 1500, 2000, 3000, 5000 and 10000 kilometers.

\begin{figure}[htbp]
\centerline{\includegraphics[width=8.7cm]{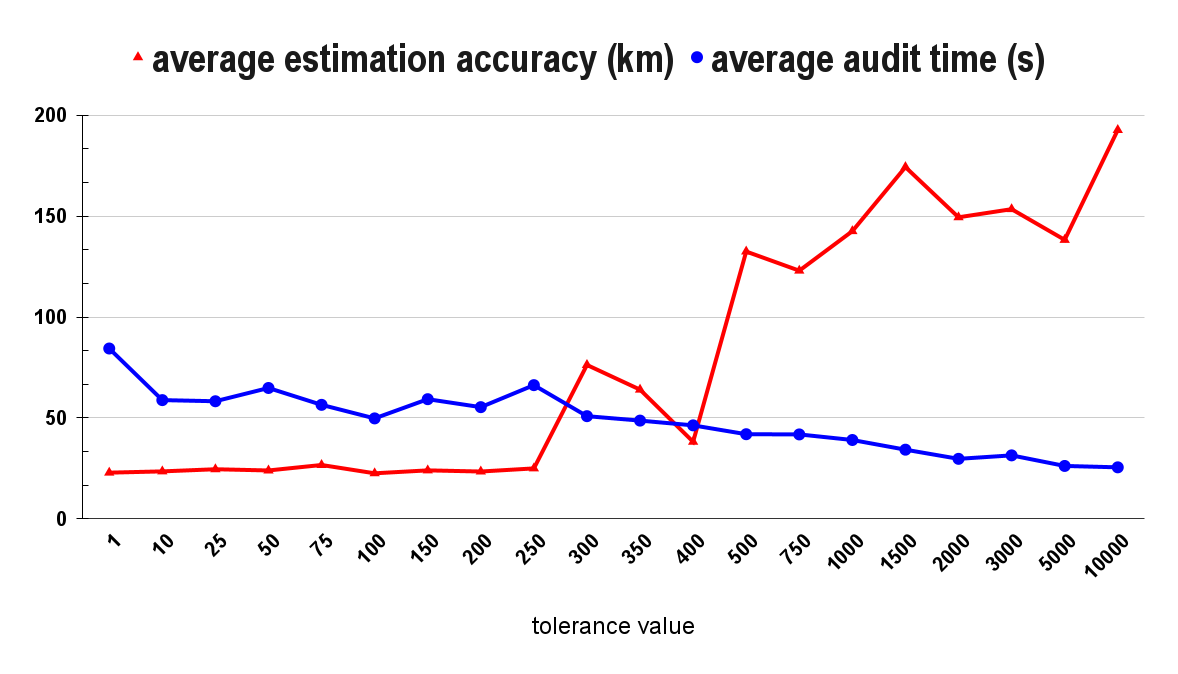}}
\caption{Average Estimation Accuracy and Average Audit Time as a Function of $Tolerance$ ($zone\_size=200$, $NB\_LM=15$ and $interval\_percent=15$)}
\label{fig:tolerance}
\end{figure}

As can be seen in Figure \ref{fig:tolerance}, lower values of $tolerance$ give satisfactory results up to $tolerance=250km$. After that, the results become very inaccurate because higher values of $tolerance$ prevent a proper correction of the estimates by leaving the main do-while loop (cf. Algorithm \ref{alg:high level alg}) too early, showing the necessity of such a correction mechanism. Likewise, audit time is reduced for higher $tolerance$ values because fewer do-while loop iterations are realized. However, very low values of $tolerance$ make the audit too long by executing superfluous do-while loop iterations that don't give more accurate results. 

In our experiment, the best result is obtained for $tolerance=100km$ with an average $dist\_estimation\_real=22.6km$ and an average audit time of $49.8s$, the lowest audit time among the satisfactory results (i.e. $tolerance<300$).

\textbf{NB\_LM. }
We set $NB\_LM$
%, the required amount of $LM_A$, 
to go from 1 to 25. From previous tests, we know that running the audit with every available landmark gives a satisfactory estimate, but is suboptimal in terms of performance.

\begin{figure}[htbp]
\centerline{\includegraphics[width=8.7cm]{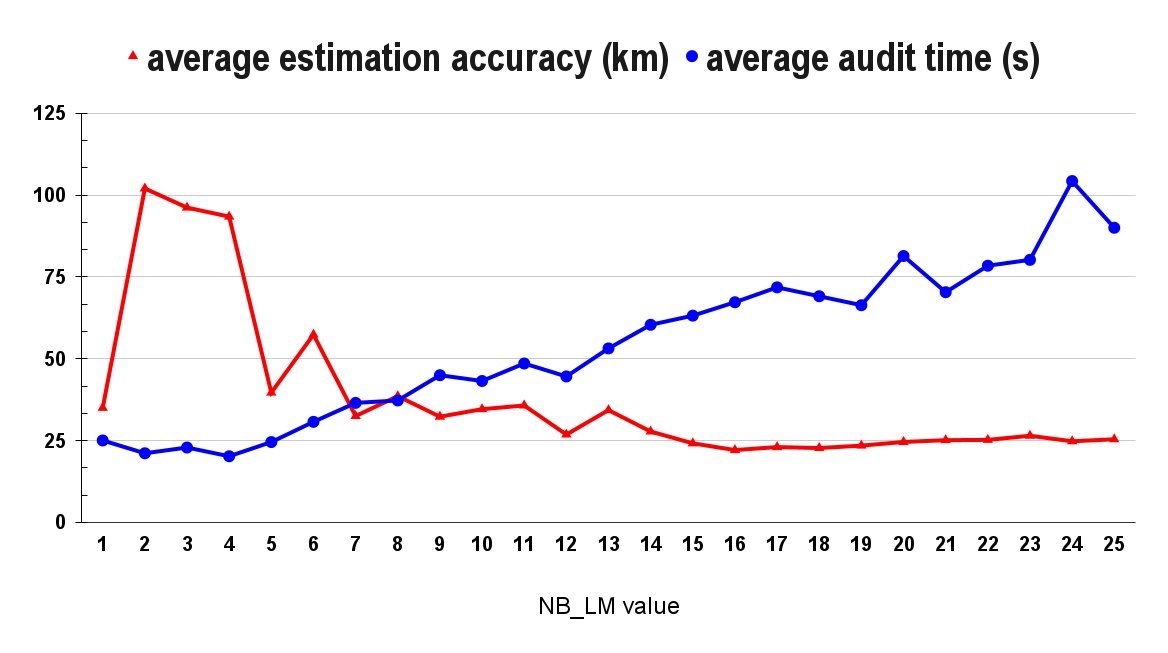}}
\caption{Average Estimation Accuracy and Average Audit Time as a Function of $NB\_LM$ ($tolerance=100$, $zone\_size=200$ and $interval\_percent=15$)}
\label{fig:nblm}
\end{figure}

As shown in Figure \ref{fig:nblm}, higher values of $NB\_LM$ give more accurate estimates than lower values, but significantly increase the total audit time, because more $LM_A$ generally means less $LM_S$. For smaller amounts of $LM_A$, the amount of $LM_S$ is too high and can't be filtered due to the small sample size of measurements, leading to less accurate estimates. On the other hand, larger amounts of $LM_A$ require more measurements to be taken and processed for the audit, resulting in performance degradation.

In our experiment, the best result is obtained for $NB\_LM=16$ with an average $dist\_estimation\_real=22.1km$ and an average audit time of $67.7s$. Although lower $NB\_LM$ are significantly faster to run and able to detect a lie in the region declaration, the estimates are too imprecise for our purposes.

\subsection{Impact of Landmark Meshing Configurations}\label{exp:landmark_meshing}

Since the available landmarks (e.g. RIPE Atlas) are unevenly distributed around the world and mostly concentrated in large urban areas, thus creating dead zones (e.g. areas without landmarks), the aim of these new experiments is to observe the influence of the landmark distribution on the accuracy and audit time of GeoFINDR. This will allow us to understand the sensitivity of the location estimation to the choice of $LM_S$, and the sensitivity of the associated DDR to the proximity of $LM_S$ to the cloud instance. 

\begin{figure}
\begin{subfigure}[t]{.475\textwidth}
  \centering
  \includegraphics[width=\linewidth]{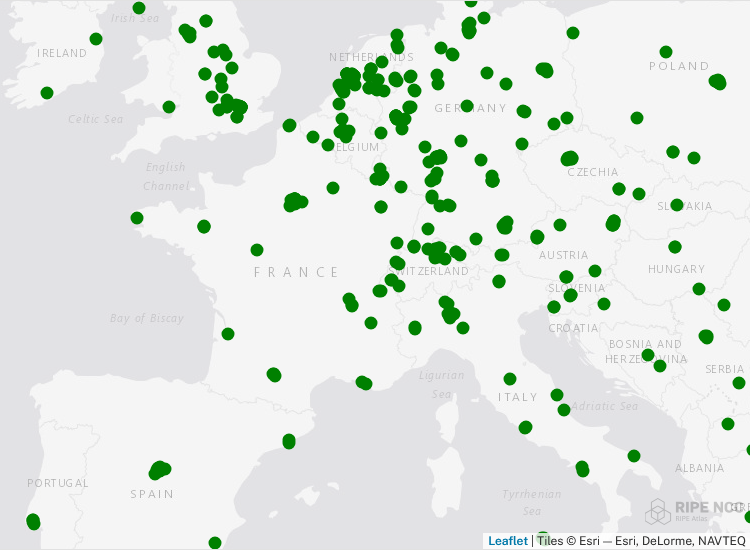}
  \caption{Mesh Configuration 1 - "Paris as a Landmark-Dense Zone", with all RIPE Atlas Western Europe Landmarks} %https://atlas.ripe.net/anchors/map
  \label{fig:meshparis}
\end{subfigure}
\hfill
\begin{subfigure}[t]{.475\textwidth}
  \centering
  \includegraphics[width=\linewidth]{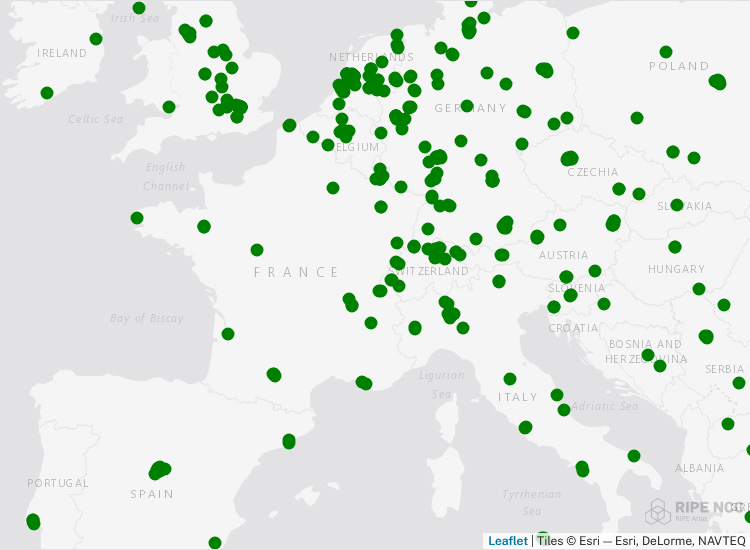}
  \caption{Mesh Configuration 2 - "Paris as a Dead Zone", where Paris Area Landmarks are Excluded from the RIPE Atlas Western Europe Landmarks List}
  \label{fig:meshwithoutparis}
\end{subfigure}
\caption{Mesh configurations with RIPE Atlas Western Europe Landmarks}
\label{fig:landmarks meshing}
\end{figure}

In previous Section \ref{exp:setup params}, experiments are conducted with nearby, densely populated landmarks in the Paris area, with the landmark distribution shown in Figure \ref{fig:meshparis}, and, with the right set of parameters, GeoFINDR is shown to be able to estimate the true position of our machine, even in the case of incorrectly declared VM locations. 
In this section, we compare the results of GeoFINDR's ability to detect a lie, based on the set of declared false positions listed in Section \ref{experimentation:protocol}, and to estimate the true location of an instance by considering two mesh configurations. The first mesh configuration considers the RIPE Atlas landmarks, including the landmarks of the Paris region (cf. Figure \ref{fig:meshparis}) and the second one simulates the Paris area as a dead zone, by removing the landmarks in the Paris area from the list of available RIPE Atlas landmarks, as shown in Figure \ref{fig:meshwithoutparis}. By removing all landmarks in the Paris area, our closest landmarks to the true Evry VM location are in the cities of Reims and Tours, which are respectively $135km$ and $185km$ away from our VM's position.

\textbf{With landmarks in the Paris area}, we obtain satisfactory estimations with an accuracy of $22.1km$ for $34.3s$ with a true declaration and an average of $22.4km$ for an average of $87.4s$ from the set of false declared positions, as shown in Table \ref{tab:mesh experiment}. In presence of multiple nearby landmarks, GeoFINDR is able to estimate the true position of the VM and thus reliably detect a false declared position.

\textbf{Without landmarks in the Paris area}, as expected, results are less accurate and audit times are longer, as shown in Table \ref{tab:mesh experiment}. We get an accuracy of $303.8km$ for $396.9s$ with a true declaration and an average accuracy of $164.3km$ for an average audit time of $244.1s$ with false declarations. Out of the 24 conducted audits, only 4 are satisfactory (accuracy below $50km$) and identify the Paris area as the true location of our instance, and the worst result achieved an accuracy of $421.8km$.\\ 
Even without landmarks in the Paris area, GeoFINDR is able to detect a lie in position declarations. In fact, we can compare the average lie extent, i.e. the difference between the falsely declared position and the true one, with the lie estimation, i.e. the distance between the declared position and the position estimate. The average lie extent in our experiment is of $3406.9km$ and we have an average lie estimation of $3445.3km$, which is only a small difference of 1.1\%. These results show that without nearby landmarks, GeoFINDR is still able to locate our instance in Western Europe, which corresponds to a continental scale estimate.  \\
The drop in performance is thus explained by the inappropriate setup parameters used for continental level estimation. The setup parameters used aim to estimate the location of our instance at a city level, but the lack of nearby landmarks changes the scale of GeoFINDR estimation accuracy. This causes the estimation to make a lot of measurements and do-while loop iterations because the solution cannot properly converge to a satisfactory solution. 

This experiment demonstrates the impact of landmark distribution and the need to have nearby landmarks to get an accurate estimation. Thus, there is a need to correctly define the scale to which GeoFINDR can reliably estimate an instance's geolocation. Therefore, we recommend to set the $tolerance$ parameter value to correspond to the size of the available nearby landmarks scale to avoid long audit times.

\begin{comment}
\begin{figure}[htbp]
\centerline{\includegraphics[width=9cm]{map_with_paris.png}}
\caption{Mesh Configuration 1 - "Paris as a Landmark-Dense Zone", with all RIPE Atlas Western Europe Landmarks} %https://atlas.ripe.net/anchors/map
\label{fig:meshparis}
\end{figure}

\begin{figure}[htbp]
\centerline{\includegraphics[width=9cm]{map_without_paris.png}}
\caption{Mesh Configuration 2 - "Paris as a Dead Zone", where Paris Area Landmarks are Excluded from the RIPE Atlas Western Europe Landmarks List}
\label{fig:meshwithoutparis}
\end{figure}
\end{comment}

\begin{table*}[htbp]
\caption{Landmark Meshing Experiment Results} \label{tab:mesh experiment}
Setup parameters: $tolerance=100$, $zone\_size=1000$, $NB\_LM=16$ and $interval\_percent=35$ 
%    \centering

    \begin{center}
    %\begin{tabularx}{\textwidth}{|>{\centering\arraybackslash}X|>{\centering\arraybackslash}X|>{\centering\arraybackslash}X|>{\centering\arraybackslash}X|}
    %\begin{tabularx}{\textwidth}{|c|c|c|c|c|c|c|}
    \begin{adjustwidth}{-0.65cm}{}
    \begin{tabularx}{18.665cm}{|c|c|c|c|c|c|c|c|}
%    \begin{tabular}{|c|c|c|c|}
        \hline
        \multicolumn{2}{|c|}{}&\multicolumn{3}{c|}{\textbf{With landmarks in Paris area}}&\multicolumn{3}{c|}{\textbf{Without landmarks in Paris area}}\\
        \hline
        
        %\textbf{Declared positions} & \textbf{Lie extent$^{\mathrm{b}}$} & Lie estimation$^{\mathrm{c}}$  & Accuracy & Audit time & Lie estimation$^{\mathrm{c}}$  & Accuracy & Audit time  \\
        \makecell{\textbf{Declared}\\ \textbf{positions}} & \textbf{Lie extent$^{\mathrm{b}}$} & \makecell{Lie \\estimation$^{\mathrm{c}}$}  & Accuracy & Audit time & \makecell{Lie \\estimation$^{\mathrm{c}}$}  & Accuracy & Audit time  \\
        \hline
        
        TRUE & 0km & 22.1km & 22.1km & 34.3s & 303.8km & 303.8km & 396.9s \\
        \hline
        
        FALSE$^{\mathrm{a}}$ & \makecell{3406.9km\\$\sigma=4319.7km$} & \makecell{3404.4km\\$\sigma=4320.3km$} & \makecell{22.4km\\$\sigma=1.9km$} & \makecell{87.4s\\$\sigma=34.4km$} & \makecell{3445.3km\\$\sigma=4300.8km$} & \makecell{164.3km\\$\sigma1=10km$} & \makecell{244.1s\\$\sigma1=43.9$km} \\
        \hline
        
    \end{tabularx}
    \end{adjustwidth}
    
    \end{center}
    \footnotesize{$^{\mathrm{a}}$Values are averages calculated on the 23 false declarations only, $\sigma$ is for the standard deviation; }
    \footnotesize{$^{\mathrm{b}}$$distance\_real\_declared$; } 
    \footnotesize{$^{\mathrm{c}}$$distance\_estimated\_declared$;} 
\end{table*}

\section{Conclusion}
\label{chap:Conclusion}
%Cloud instance geolocation is an important challenge to ensure legal, technical and economic aspects for customers using a highly distributed architecture such as multicloud. 
Our proposed GeoFINDR approach is a practical software approach to geolocate multiple cloud instances by developing a program that runs from each cloud instance to determine its true location individually. 
To do this, we redefined and corrected several weaknesses in the key steps of similar work. We redefined the threat model to be more realistic with respect to the actual capabilities of a cloud service provider. We defined a two-step selection of landmark sets $LM_A$ and $LM_S$. The selection of $LM_A$ is performed by an original Greedy algorithm of dispersed points selection called Dispoints. RIPE Atlas anchors are used as landmarks for their real-time open source measurements and worldwide distribution. From $LM_A$, we find a set of $LM_S$ to derive the distance-delay relation and estimate the position of the instance using a method that calculates the barycenter of $LM_S$ to always determine a position.\\
By using the latest measurements from RIPE Atlas, the barycenter computation to derive the estimated position, and a do-while correction loop, GeoFINDR is able to reliably detect a lie in the CSP's declaration and obtain an average accuracy of $22.6km$ from multiple declared positions around the world. 
GeoFINDR's accuracy and audit time depends on a set of parameter values and the distribution of landmarks in the area of the instance's physical location. With good settings, GeoFINDR estimates the position of an instance in less than two minutes, which meets the need to verify the location of instances, to meet legal or security requirements. 
In the event of a fraud in the instance's declared position, GeoFINDR provides all the data the customer needs to investigate, using in-cloud delays, quality metrics and estimated distances between declared and estimated positions.

GeoFINDR is a solution suitable for any distributed infrastructure that requires its cloud instances to be geolocated at a city or country scale, and where the required responsiveness meets current audit times. 
The source code and experiment results and outputs are available in the public GeoFINDR GitHub repository to ensure its continuity.
However, further work is needed to evaluate the GeoFINDR capabilities with other fraud methods (e.g. use of dark fibers, VPN tunnel or tempering), to experiment from different real-world locations and infrastructures, and to reduce the audit time by multithreading the measurements to make our solution more robust and adapted to highly dynamic deployments. 

The ability of customers to locate their resources in the cloud with reliability and transparency from CSPs still faces many challenges. However, we believe that GeoFINDR is a step towards building a cloud of trust.

\end{document}